\documentclass[12pt]{article}

\usepackage{scicite}
\usepackage{amsmath,amssymb,graphicx,textcomp}
\usepackage{color}

\usepackage{times}

\newenvironment{sciabstract}{%
\begin{quote} \bf}
{\end{quote}}

\newcounter{lastnote}

\title{Resonances arising from hydrodynamic memory in Brownian motion - The colour of thermal noise}

\author
{Thomas Franosch,$^{1,\text{\S}}$ Matthias Grimm,$^{2,3,\text{\S}}$ Maxim Belushkin,$^{4}$ Flavio Mor,$^{3}$\\
Giuseppe Foffi,$^{4}$ L\'{a}szl\'{o} Forr\'{o},$^{3}$ Sylvia Jeney$^{2,3,\ast}$\\
\\
\normalsize{$^{1}$Institut f\"ur Theoretische Physik, Universit\"at Erlangen-N\"urnberg,}\\
\normalsize{Staudtstra{\ss}e 7, 91058 Erlangen, Germany}\\
\normalsize{$^{2}$M. E. M\"{u}ller Institute for Structural Biology, Biozentrum,}\\
\normalsize{University of Basel, Klingelbergstrasse 70, Basel 4056, Switzerland}\\
\normalsize{$^{3}$Laboratoire of Physics of Complex Matter,}\\
\normalsize{Ecole Polytechnique F\'ed\'erale de Lausanne (EPFL), CH-1015 Lausanne, Switzerland}\\
\normalsize{$^{4}$Institute of Theoretical Physics,}\\
\normalsize{Ecole Polytechnique F\'{e}d\'{e}rale de Lausanne (EPFL), CH-1015 Lausanne, Switzerland}\\
\\
\normalsize{$^{\text{\S}}$These authors contributed equally to this work.}\\
\normalsize{$^\ast$To whom correspondence should be addressed;}
\normalsize{E-mail: sylvia.jeney@epfl.ch}
}

\normalsize{\date{\today}}

\begin{document}

\baselineskip24pt

\maketitle

\begin{sciabstract}

\textbf{Observation of the Brownian motion of a small probe interacting with its environment is one of the main strategies to characterize soft matter$^{\text{1-4}}$. Essentially two counteracting forces govern the motion of the Brownian particle. First, the particle is driven by the rapid collisions with the surrounding solvent molecules, referred to as thermal noise. Second, the friction between the particle and the viscous solvent damps its motion. Conventionally, the thermal force is assumed to be random and characterized by a white noise spectrum. Friction is assumed to be given by the Stokes drag, implying that motion is overdamped. However, as the particle receives momentum from the fluctuating fluid molecules, it also displaces the fluid in its immediate vicinity. The entrained fluid acts back on the sphere and gives rise to long-range correlation$^{\text{5,6}}$. This hydrodynamic memory translates to thermal forces, which display a coloured noise spectrum. Even 100 years after Perrin�s pioneering experiments on Brownian motion$^{\text{7}}$, direct experimental observation of this colour has remained elusive$^{\text{8}}$. Here, we measure the spectrum of thermal noise by confining the Brownian fluctuations of a microsphere by a strong optical trap. We show that due to hydrodynamic correlations the power spectral density of the spheres positional fluctuations exhibits a resonant peak in strong contrast to overdamped systems. Furthermore, we demonstrate that peak amplification can be achieved through parametric excitation. In analogy to Microcantilever-based sensors$^{\text{9,10}}$ our results demonstrate that the particle-fluid-trap system can be considered as a nanomechanical resonator, where the intrinsic hydrodynamic backflow enhances resonance. Therefore, instead of being a disturbance, details in thermal noise can be exploited for the development of new types of sensors and particle-based assays for lab-on-a-chip applications$^{\text{11,12}}$.}

\end{sciabstract}

Einstein�s theory of Brownian motion published in 1905$^{\text{13}}$ received considerable attention, and was later reformulated in terms of a Langevin equation$^{\text{14}}$. There, particle motion is driven by thermal fluctuations induced through collisions with the fluid molecules. These rapid kicks are assumed to be random and independent at frequencies much smaller than the collision rate of $\sim 1\,\text{THz}$. The thermal force spectrum is then constant over a wide range of frequencies, which is commonly referred to as white noise spectrum$^{\text{14}}$.
However, when the densities of particle and fluid are comparable, their coupling becomes important$^{\text{15,16}}$. As the suspended particle fluctuates through the solvent, long-range correlations build up due to momentum exchange, leading to hydrodynamic memory in the solvent by the particle's past motion.
By the fluctuation-dissipation theorem, the statistics of the thermal force is characterized through a delta-correlated white-noise term, and a coloured frequency-dependent component, which reflects the retarded viscous response of the fluid continuum to the particle.

To directly measure the predicted correlations in thermal noise, we combined strong optical trapping with high-resolution interferometric detection$^{\text{17}}$. Under such conditions, the force balance for the sphere, ignoring its inertia, reads $F_\text{fr}(t) -K x(t) + F_\text{th}(t)=0$, where $x(t)$ is the bead's displacement from the trap center, $K$ the stiffness of the harmonic optical trap, and $F_\text{fr}(t)$ the friction force on the particle. At low frequencies, respectively long times, the trapping force dominates over friction, and the Langevin equation reduces to $K x(t) \simeq F_\text{th}(t)$.
Consequently, by monitoring the fluctuating motion of the bead, we effectively probe the thermal force of the fluid $^{\text{8}}$. At long times, correlations in thermal noise
\begin{equation}
\langle F_\text{th}(t) F_\text{th}(0) \rangle \simeq K^2\langle x(t) x(0) \rangle ,
\end{equation}
become directly accessible by calculating the positional autocorrelation function, $\text{PAF}(t) = \langle x(t) x(0)\rangle$ from the recorded position fluctuations of the trapped sphere (Fig. 1a). Similarly, deviations from white noise in the power spectral density (PSD) of the displacement reflect the colour of thermal force $^{\text{8}}$.

An overview of the characteristic time scales for trapped Brownian motion is displayed in Fig. 1b. Comparing Stokes friction with the harmonic restoring force suggests the trap relaxation time $\tau_\text{K} = \gamma/K$ as the longest characteristic time scale, where $\gamma = 6 \pi \eta R$ is the stationary friction coefficient of the bead for macroscopic  no-slip boundary conditions,  $\eta$ the shear viscosity of the fluid and $R$ the radius of the sphere. Momentum is transferred from the particle to the fluid at times $\tau_\text{p}=m/\gamma$, with $m$ the mass of the bead.
Based on these contributions, and for not too strong trapping, $\tau_\text{K} > 4 \tau_\text{p}$, the bead behaves like a harmonic oscillator in the overdamped regime. The corresponding PSD matches a Lorentzian, $\text{PSD}(f) = 2 (k_B T/K) \tau_\text{K}/[1 + (2\pi f \tau_\text{K})^2]$ for frequencies around the corner frequency, $f_\text{K} = 1/2\pi\tau_\text{K}$.  Furthermore, the PAF displays an exponential relaxation $\text{PAF}(t) = (k_B T/K) \exp( - t/\tau_\text{K})$ at time scales $ t \simeq \tau_\text{K}$.
However, in a fluid this picture is incomplete, as hydrodynamic backflow has to build up first by vorticity diffusion (Fig. 1b, middle), and the  additional time scale, $\tau_\text{f}= R^2/\nu$, becomes important, where $\nu$ denotes the kinematic viscosity of the fluid.
The sequential order and magnitude of the various time scales depends on the size and mass of the Brownian particle, the nature of the solvent, and the stiffness of the optical trap.

For a given solvent and particle, it is possible to investigate the dynamics of the system at different times by increasing $K$, thus shifting $\tau_\text{K}$ towards $\tau_\text{f}$ or $\tau_\text{p}$. The experimental challenge consists in exploring a wide range of time scales. Therefore, we optimized our set-up$^{\text{18}}$ to obtain high trapping strength and best signal-to-noise ratio with a bandwidth covering at least four decades. The achieved resolution was on the order of $1\text{nm}$ in space and close to $1\text{\textmu s}$ in time.

In typical optical tweezers experiments, silica or polystyrene spheres immersed in water and with sizes $\lesssim 1\text{\textmu m}$ are used. This yields values for $\tau_\text{f}$ and $\tau_\text{p}$  below $1\text{\textmu s}$, thus, below our temporal resolution limit$^{\text{19}}$. Instead, we used melamine resin beads with a density $\rho_\text{p}=1510\,\text{kg}/\text{m}^{3}$ and a radius of $1.0 \pm 0.1\text{\textmu m}$ or $1.5 \pm 0.1\text{\textmu m}$ suspended in an acetone solution with $\nu = 5 \cdot 10^{-7} \text{m}^2/\text{s}$ at $22^\circ\text{C}$. Then, $\tau_\text{p} \approx 1\,\text{\textmu s}$ and $\tau_\text{f} \approx 3\,\text{\textmu s}$.
Furthermore, the difference between the refractive indices of resin ($n=1.68$) and acetone ($n=1.36$) was high enough to provide a good trapping efficiency$^{\text{20}}$ and allowed for the visualization of the particle by optical microscopy, which was not the case for e.g. silica particles ($n=1.37$).
With such an experimental setup, the magnitude of the trap relaxation time $\tau_\text{K}$ could be brought close to $\tau_\text{f}$, but still with $\tau_\text{K} > 4 \tau_\text{p}$.

The position fluctuations $x(t)$ of a single resin sphere confined by the stiffest trap achievable with our set-up were acquired during $\mathcal{T}\approx 50\,\text{s}$ at a sampling rate of $1\,\text{MHz}$. A comparison of $\text{PSD}(f)$ at different experimental conditions (Fig. 2) confirmed that the choice of the bead size, fluid, and trap stiffness, hence $\tau_\text{f}$, respectively $\tau_\text{K}$, determined the emergence of a resonance. When using water (red  symbols), the maximal corner frequency $f_\text{K}$ we obtained resulted in an enhancement of the PSD  close to our noise limit.  Yet, striking deviations from a simple Lorentzian (red line), typical of overdamped systems, are clearly visible. The measured decay is steeper and shifted to higher frequencies. Replacing water by acetone decreased the ratio $\tau_\text{K}/\tau_\text{f}$. As a consequence, a resonance peak  appears indicating an enhanced spectrum of the thermal force with increasing frequencies. Thus, deviations from white noise are towards the blue at frequencies, which are by orders of magnitude smaller than the collision rate of the solvent molecules.
In the time domain, the corresponding $\text{PAF}(t)$ exhibits a zero crossing (Fig. 3, green and blue symbols) resulting from anticorrelations, which are in remarkable contrast to the purely relaxing behavior of overdamped motion. This negative overshoot reflects the resonant peak in the PSD. The long-time decay is significantly slower than exponential and corroborates persistent correlations. Deviation from an exponential relaxation (green and blue lines) are substantial already at short times, characterized by a slower and then more rapid decay.

For a quantitative description, we use a Langevin equation with no-slip boundary conditions accounting for slow vortex diffusion$^{\text{15,16}}$ and trapping$^{\text{21,22}}$. Our data for the PSD are in excellent agreement with the theoretical expression (black lines in Fig. 2), and were fitted simultaneously in the two dimensions lateral to the optical axis using the same $\tau_\text{f}$. We achieved trap stiffnesses of up to $345\,\text{\textmu N/m}$. The shape of the PSD is only determined by the ratio $\tau_\text{K}/\tau_\text{f}$, and hence reflects, at a given bead radius, the fluid properties through $\tau_\text{f}$, but is independent of the particle's mass.
The fitted parameters were then used to compare the measured $\text{PAF}(t)$ with theory$^{\text{22}}$. The theoretical curves (black lines in Fig. 3) capture the data quantitatively over two decades in time and three orders of magnitude in signal, highlighting our unprecedented signal to noise ratio.
We observe a hydrodynamic power-law tail $\text{PAF}(t) \simeq - (k_B T\gamma/K^2) \sqrt{\tau_\text{f}/4 \pi t^3}$, which by Eq. (1) directly reflects the corresponding persistent correlations in the thermal forces
\begin{equation}
 \langle F_\text{th}(t) F_\text{th}(0) \rangle = -k_B T \gamma  \sqrt{\tau_\text{f}/4 \pi}\ t^{-3/2} \qquad  \text{ for }  t>0\, .
\end{equation}

As just shown, experimental access to short times reveals a resonance in Brownian motion, where overdamped motion is commonly assumed. The main technical achievement is to strongly confine the particle and to detect its position fluctuations at highest bandwidth$^{\text{8}}$. For applications, it is desirable to enhance this effect. Stronger and narrower resonances can be obtained in the inertial regime, where Brownian motion is also sensitive to the particle's mass$^{\text{23,24}}$. To reach this window, $\tau_\text{K}$ has to be approached to $\tau_\text{p}$ by increasing $K$ or $m$.
Although heavier particles, which simultaneously allow for more efficient trapping are still to be developed$^{\text{25}}$, the timeline displayed in Fig. 1b can be explored theoretically and by means of computer simulations.
The appropriate Langevin equation, $m\ddot{x}(t)= F_\text{fr}(t)- K x(t) + F_\text{th}(t)$, comprises then a mass term, which eventually becomes larger than friction. By decreasing $\tau_\text{K}$, the resonance is increased (Fig. 4a, continuous lines). Interestingly, compared to the simple overdamped harmonic oscillator (dashed lines), the peak is significantly enhanced and its position, $f_\text{max}$, is shifted to lower frequencies by the contribution of hydrodynamic memory.

The study of Brownian motion at short times in a medium displaying hydrodynamic effects has also become accessible by advanced simulation techniques. We employed multi-particle collision dynamics$^{\text{26}}$ with a molecular dynamics coupling between the solute and solvent particles. The method yielded a highly compressible solvent and reflected correctly the hydrodynamic effects at coarse-grained scales.
The approach was implemented most conveniently for full-slip boundary conditions at the solute-solvent interface, whereas our experiments obeyed no-slip boundary conditions. The results of the PSD corresponding as closely as possible to our macroscopic system show that a resonance emerged in the friction-dominated, as well as in the inertial regime (Fig. 4b, full circles). However, weak coupling between the Brownian particle and the surrounding fluid yielded a much weaker resonance as opposed to no-slip conditions (Fig. 4a). The collected data are in semi-quantitative agreement with the theoretical curves (Fig. 4b, coloured lines), where the friction coefficient is evaluated for a compressible fluid with full-slip boundary conditions$^{\text{27}}$.

Further peak amplification can be achieved via parametric resonance$^{\text{24,28}}$ by periodically modulating the trap strength, $K(t) = K [1 + g \cos(2 \pi f_\text{exc} t)]$. We obtained the theoretical $\text{PSD}_\text{exc}(f)$ normalized to the initial value of the non-excited PSD from the solution of a parametrically modulated Langevin equation, including hydrodynamic memory, by second order perturbation theory in the reduced modulation amplitude $g$. Similar to the harmonic oscillator, also in the presence of coloured friction best excitation is achieved at a frequency $f_\text{exc}\simeq 2f_\text{max}$, yielding an increase of up to 20\% when $g=50\%$ (Fig. 4c, empty circles). Comparable results were obtained within computer simulations (empty circles in Fig. 4b).

In view to develop particle-based assays$^{\text{11,12}}$, we evaluated the sensitivity of the peak by slightly changing the size of the particle. As shown in the inset of Fig. 4c, a change of 5\% in the bead radius results in a resonance frequency shift of 1 kHz.  We anticipate that changes in the particle's morphology, like swelling or a reaction occurring at the surface of such a Brownian probe will alter its short-time dynamics and become detectable. As also single cells, microorganisms and microcarriers can be bound harmonically $^{\text{10,29}}$, short-time detection of their Brownian fluctuations may become a sensitive way to characterize their state or evolution in native solutions and without specific markers. Reciprocally, also changes in the medium surrounding the probing particle modulate its fluctuation spectrum$^{\text{2-4}}$, and a dynamic polymer system may be studied in great detail.\\

\textbf{METHODS SUMMARY}

Melamine resin spheres (Sigma-Aldrich, Analytix 5, 2001) were suspended in high purity acetone at minimal concentrations to allow trapping and observation of exclusively one single particle. The sample chamber was made of a microscope slide having a ground-in polished cavity in its center ($18\text{mm}$ in diameter and $0.8\text{mm}$ in depth) (Marienfeld GmbH, Germany). The cavity was filled with the bead-acetone mixture and sealed with a glass coverslip (thickness $\sim\! 130\text{\textmu m}$) and high viscosity vacuum grease to prevent evaporation. Special care was taken not to introduce bubbles into the sample chamber. After loading, the sample chamber was mounted onto our custom-made inverted microscope-optical trap set-up.

Since the trap was asymmetric$^{\text{20}}$, we obtained different $\tau_\text{K}$ for each dimension from fitting our data to the theory. Both values differed by maximally $20\%$. The determined values for $\tau_\text{f}$ and $\tau_\text{K}$ were used to calculate $\nu$ and $K$. The bead radii were chosen to be within the tolerances given by the manufacturer. The slightly higher acetone viscosities of $\nu = 5 \cdot 10^{-7} \text{m}^2/\text{s}$  at the laboratory temperature of $22^\circ\text{C}$, compared to the literature value of $\nu = 4 \cdot 10^{-7} \text{m}^2/\text{s}$ were attributed to a contamination of our acetone stock solution by atmospheric water during experimentation. {\it In situ} viscometry measurements$^{\text{30}}$ were performed to cross-check for the value of the acetone viscosity obtained through the determination of $\tau_\text{f}$. We obtained $\nu = (4.8 \pm 0.6) \cdot 10^{-7} \text{m}^2/\text{s}$ for 10 measurements with different beads. As acetone has a very low absorption coefficient at our laser wavelength $\lambda=1064\,\text{nm}$, no heating of the solution was expected.\\

\textbf{Full Methods} and any associated references are available at the end of the manuscript.

\newpage

\textbf{REFERENCES}
\begin{itemize}
\item[1.]
Mason, T. G. \& Weitz, D. A. Optical measurements of frequency-dependent linear viscoelastic moduli of complex fluids. {\it Phys. Rev. Lett.} {\bf74}, 12501253 (1995).
\item[2.]
Nowak, A. P. \emph{et al.} Rapidly recovering hydrogel scaffolds from self-assembling diblock copolypeptide amphiphiles. \emph{Nature} {\bf417}, 424-428 (2002).
\item[3.]
Gardel, M. L. \emph{et al.} Elastic behavior of cross-linked and bundled actin networks. {\it Science} {\bf304}, 13011305 (2004).
\item[4.]
Chaudhuri, O., Parekh, S. H., Fletcher, D. A. Reversible stress softening of actin networks. \emph{Nature} {\bf445}, 295-298 (2007).
\item[5.]
Alder, B. J. \& Wainwright, T. E. Velocity autocorrelations for hard spheres. {\it Phys. Rev. Lett.} {\bf18}, 988-990 (1967).
\item[6.]
Jeney, S., Luki\'c, B., Kraus, J. A., Franosch, T. \& Forr\'o, L. Anisotropic memory effects in confined colloidal diffusion. {\it Phys. Rev. Lett.} {\bf100}, 240604 (2008).
\item[7.]
Perrin, J. Brownian movement and molecular reality. Translated by F. Soddy (Taylor and Francis, London, 1910). The original paper, Le Mouvement Brownien et la R\'{e}alit\'{e} Mol\'{e}culaire, appeared in the Ann. Chim. Phys. {\bf18}, 5�114 (1909).
\item[8.]
Berg-S{\o}rensen K. \& Flyvbjerg H. The colour of thermal noise in classical Brownian motion: a feasibility study of direct experimental observation. {\it New J. of Phys.} {\bf7}, 38 (2005).
\item[9.]
Fritz, J. \emph{et al.} Translating biomolecular recognition into nanomechanics. {\it Science} {\bf288}, 316-318 (2000).
\item[10.]
Burg, T. P. \emph{et al.} Weighing of biomolecules, single cells and single nanoparticles in fluid. {\it Nature} {\bf446}, 1066-1069 (2007).
\item[11.]
Braeckmans, K., De Smedt, S. C. , Leblans, M., Pauwels, R., Demeester, J. Encoding microcarriers: present and future technologies. \emph{Nature Reviews Drug Discovery} \textbf{1}, 447-456 (2002).
\item[12.]
Craighead H. Future lab-on-a-chip technologies for interrogating individual molecules. \emph{Nature} \textbf{442}, 387-393 (2006).
\item[13.]
Einstein, A. \"{U}ber die von der molekularkinetischen Theorie der W\"{a}rme geforderte Bewegung von in ruhenden Fl\"{u}ssigkeiten suspendierten Teilchen. {\it Ann. d. Phys.} {\bf17}, 549-560 (1905).
\item[14.]
Ming Chen Wang \& Uhlenbeck, G. E. On the theory of the Brownian motion II. {\it Rev. Mod. Phys.} {\bf 17}, 323-342 (1945).
\item[15.]
Vladimirsky, V. \& Terletzky, Y. A. Hydrodynamical Theory of Translational Brownian Motion. {\it Zh. Eksp. Theor. Fiz.} {\bf15}, 258-263 (1945). For an English discussion see V. Lisy and J. Tothova, arXiv:cond-mat/0410222v1.
\item[16.]
Hinch, E. J. Application of the Langevin equation to fluid suspensions. {\it J. Fluid Mech.} {\bf72}, 499-511 (1975).
\item[17.]
Gittes, F. \& Schmidt, C. F. Interference model for back-focal-plane displacement detection in optical tweezers. {\it Opt. Lett.} {\bf23}, 7-9 (1998).
\item[18.]
Jeney, S., Mor, F., K\H{o}szali, R., Forr\'{o}, L., Moy, V. T. Monitoring ligand-receptor interactions
by photonic force microscopy. {\it Nanotechnology} {\bf21}, 255102 (2010).
\item[19.]
Luki\'{c}, B. \emph{et al.} Motion of a colloidal particle in an optical trap. {\it Phys. Rev. E} {\bf76}, 011112 (2007).
\item[20.]
Rohrbach, A. Stiffness of optical traps: quantitative agreement between experiment and electromagnetic theory. {\it Phys. Rev. Lett.} 95, 168102 (2005).
\item[21.]
Berg-S{\o}rensen, K. \& Flyvbjerg, H. Power spectrum analysis for optical tweezers. {\it Rev. Sci. Inst.} {\bf75}, 594-612 (2004).
\item[22.]
Clercx, H. J. H. \& Schram, P. P. J. M. Brownian particles in shear flow and harmonic potentials: A study of long-time tails. {\it Phys. Rev. A} {\bf46}, 1942-1950 (1992).
\item[23.]
Huang, R. \emph{et al.} Direct observation of ballistic Brownian motion of a single particle in a liquid. Under review (2010).
\item[24.]
Di Leonardo, R. \emph{et al.} Parametric Resonance of Optically Trapped Aerosols. {\it Phys. Rev. Lett.} {\bf99}, 010601 (2007).
\item[25.]
Bormuth, V. \emph{et al.} Optical trapping of coated microspheres. {\it Opt. Express} {\bf16}, 13831-13844 (2008).
\item[26.]
Padding, J. T. \& Louis, A. A. Hydrodynamic interactions and Brownian forces in colloidal suspensions: Coarse-graining over time and length scales. {\it Phys. Rev. E} {\bf74}, 031402 (2006).
\item[27.]
Erba{\c{s}}, A., Podgornik, R., Netz, R. R. Viscous compressible hydrodynamics at planes, spheres and
cylinders with finite surface slip. {\it Eur. Phys. J. E} {\bf 32}, 147-164 (2010).
\item[28.]
Pedersen, L. \& Flyvbjerg, H. Comment on "Direct measurement of the oscillation frequency in an optical-tweezers trap by parametric excitation". {\it Phys. Rev. Lett.} {\bf98}, 189801 (2007).
\item[29.]
Ashkin, A., Dziedzic, J. M., Yamane, T. Optical trapping and manipulation of single cells using infrared-laser beams. {\it Nature} {\bf330}, 769-771 (1987).
\item[30.]
Guzm\'{a}n, C. \emph{et al.} In situ viscometry by optical trapping interferometry. {\it Appl. Phys. Lett.} {\bf93} 184102 (2008).
\end{itemize}

\textbf{Acknowledgments} S.J. acknowledges the Swiss National Science Foundation (grant no. 200021-113529 and 206021-121396). M.G. is supported by  the NCCR for nanoscale science and the German Academic Exchange Service (DAAD). M.B. and G.F. acknowledge support by the Swiss National Science Foundation (grant no. 200021-105382/1 and PP0022 119006). We thank R. K\H{o}szali for technical help and E. Sackmann, B. U. Felderhof, U. Aebi, H. Flyvbjerg, S. Melchionna for discussions.

\textbf{Author Contributions} T.F. and M.G. contributed to the planning of the experiments, designed parts of the data analysis software, fitted the theory to the data, and interpreted the data. M.B. contributed to the fitting of the data, performed the simulations. F.M. contributed to the optimization of the experimental set-up. G.F. contributed to the simulations. L.F. contributed to the planning of the experiments. S.J. constructed and characterized the experimental set-up, conceived planned and carried out the experiments, designed the data analysis software, interpreted the data. All authors contributed, discussed and commented on the manuscript.

\textbf{Author Information} The authors declare no competing financial interests.
\\

\newpage

\subsection*{FIGURES AND LEGENDS}

\includegraphics[width=0.8\linewidth]{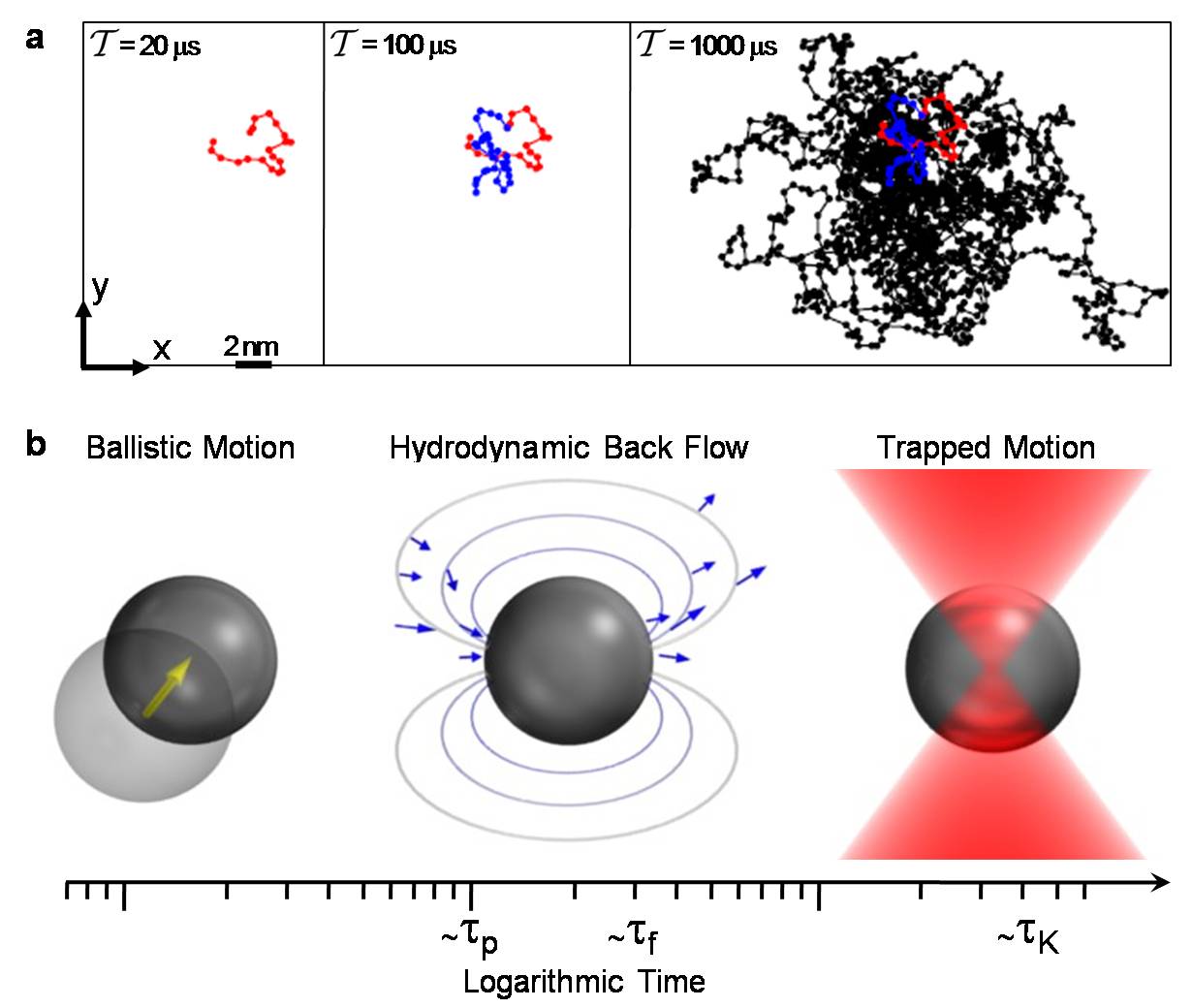}\\
\textbf{Figure 1 $\mid$ Characteristic time scales for a Brownian particle confined by the harmonic potential of an optical trap. a,} Trajectories of the trapped sphere measured in the 2 dimensions lateral to the optical axis, at three different time intervals. \textbf{b,} At very short times $t\ll\tau_\text{p}$ the particle undergoes ballistic motion (left). At the hydrodynamic time scale $\tau_\text{f}$ hydrodynamic backflow develops (center; solid lines show the emerging fluid velocity field, arrows are obtained as part of the present computer simulations). Finally, for $t\geq\tau_\text{K}$ the harmonic potential of the optical trap sets in and confines the diffusion of the particle (right).\\
\\
\newpage
\includegraphics[width=0.7\linewidth]{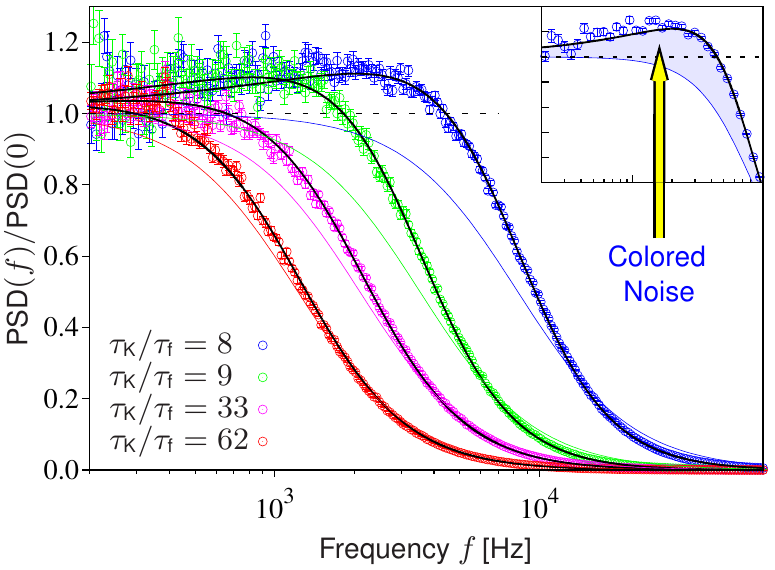}\\
\textbf{Figure 2 $\mid$ Emergence of a coloured peak in the thermal noise spectrum.} Log-linear plot of the $\text{PSD}$ normalized to its zero-frequency value $2 k_B T \tau_\text{K}/K$ of a 
melamine resin sphere in pure water 
 (red: $\tau_\text{f}=2.1 \text{\textmu s}$, $\tau_\text{K}=130\text{\textmu s}$, $R=1.42\,\text{\textmu m}$, $\nu = 9 \cdot 10^{-7} \text{m}^2/\text{s}$, $k=196\,\text{\textmu N/m}$) 
or in acetone solution ($\nu = 5 \cdot 10^{-7} \text{m}^2/\text{s}$) with increasing $\tau_\text{f}/\tau_\text{K}$ ratios  
(magenta: $\tau_\text{f}= 2.3\text{\textmu s}$, $\tau_\text{K}=75.9\text{\textmu s}$, $R=0.96\,\text{\textmu m}$, $K=77\,\text{\textmu N/m}$; green: $\tau_\text{f}=5\text{\textmu s}$, $\tau_\text{K}=45\text{\textmu s}$, $R=1.42\text{\textmu m}$, $K=190\,\text{\textmu N/m}$; blue: $\tau_\text{f}= 2.4\text{\textmu s}$, $\tau_\text{K}=19.2\text{\textmu s}$, $R=0.99\,\text{\textmu m}$, $K=309\,\text{\textmu N/m}$).
 Data were collected for $\mathcal{T}\approx 50\,\text{s}$, corresponding to $\sim5 \cdot 10^7$ datapoints. Error bars indicate the standard error on the mean.
The PSD were computed from overlapping windows of $2^{22}$ points, blocked in $100$ bins per decade. 
The black lines correspond to the hydrodynamic theory$^{\text{21}}$. The parameters $\tau_\text{f}$ and $\tau_\text{K}$ were extracted from the fit. 
The coloured lines are  Lorentzian spectra corresponding to non-hydrodynamic systems. Inset: Zoom into the enhancement of the PSD, which reflects the colour of thermal
 noise ($\tau_\text{K}/\tau_\text{f}= 8$, blocked in $20$ bins per decade).\\
\\
\newpage
\includegraphics[width=0.7\linewidth]{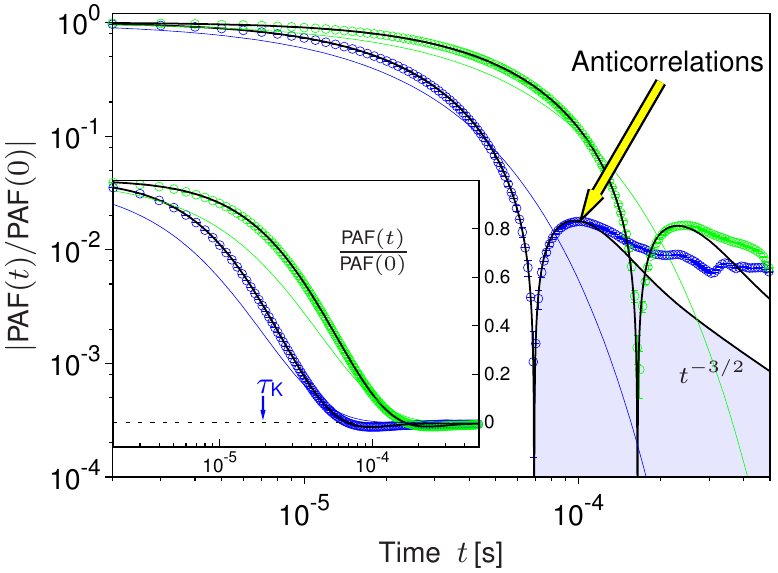}\\
\textbf{Figure 3 $\mid$ Anticorrelations in the PAF reveal that Brownian motion is not overdamped.} Double logarithmic representation of the normalized PAF(t) for two datasets of Fig. 2 (blue: $\tau_\text{K}/\tau_\text{f}=8$, green: $\tau_\text{K}/\tau_\text{f}=9$). The black lines correspond to the full hydrodynamic theory including inertial effects$^{\text{22}}$. The persistent anticorrelations are visible after the zero crossing (narrow spike) and approach a $t^{-3/2}$ power law decay. The blue, respectively green lines indicate exponential relaxations and serve as guide to the eye. Inset: Semilogarithmic plot of the same data.\\
\\
\newpage
\includegraphics[width=0.85\linewidth]{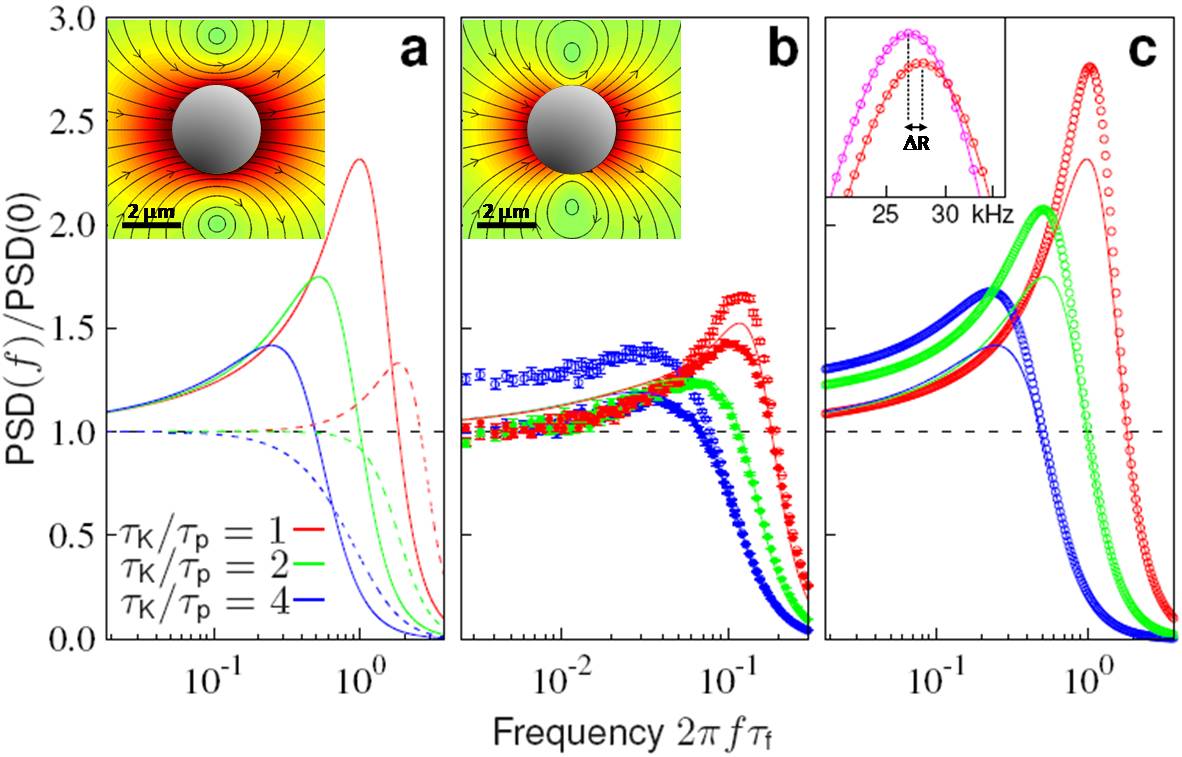}\\
\textbf{Figure 4 $\mid$ Amplification of the resonant peak. a,} Theoretical PSD$^{\text{21}}$  for a resin sphere in acetone ($R=1.5\text{\textmu m}$, $\tau_\text{p}=2.3\text{\textmu s}$, $\tau_\text{f}=5.4\text{\textmu s}$) for very strong traps ($K=1.02\,\text{, }2.04\,\text{, }4.07\,\text{mN/m}$; blue, green, and red graphs, respectively). The dashed lines correspond to the damped harmonic oscillator$^{\text{14}}$. Inset; Corresponding plot of the fluid's velocity field developing around the sphere moving along the x-direction for no-slip boundary conditions at $f=1/(2\pi\tau_{\text{f}})$. The arrows indicate the direction of the velocity field, while the colours denote the velocity magnitude rising from green to red. \textbf{b,} Simulation data (full circles) for an equivalent system with $\tau_\text{p}=0.67\tau_\text{f}$ evaluated in a compressible fluid for full-slip boundary conditions, $\gamma = 4\pi \eta a$. The coloured theoretical lines account for vortex diffusion as well as for sound waves$^{\text{27}}$. The empty circles are simulation data for an excitation with $g=50\%$ and $f_\text{exc} =2 f_\text{max}$. Inset; Corresponding velocity field at $f=1/(2\pi\tau_{\text{f}})$. \textbf{c,} $\text{PSD}_\text{exc}(f)$ for the same particle as in \textbf{a} with $g= 50\%$ and $f_\text{exc}\simeq2f_\text{max}$ (empty circles). Inset; Shift of the peak frequency when increasing the radius of the particle by $5\%$ respectively $75\text{nm}$.
All curves are normalized to the zero-frequency value of the respective non-excited PSD.

\newpage

\textbf{FULL METHODS}

\textbf{Instrument Set-up.} Our Instrument consists of a custom-built inverted light microscope with a three-dimensional (3D) sample positioning stage, an infrared laser for optical trapping, and a quadrant photodiode for high-resolution and time-resolved position detection.  The trapping beam is produced by a diode-pumped, ultra-low noise Nd:YAG laser with a wavelength of $\lambda =1064$nm (IRCL-500-1064-S, CrystaLaser, USA), and a maximal light power of $500$mW in continuous wave mode. Best trapping efficiency is achieved by expanding the effective laser beam diameter 10 times with a telecentric lens system (Sill Optics, Germany). Further details on the set-up are given in$^{\text{18}}$ and its online supplement.

\textbf{Data analysis.} Since the location of the trap center is not known with high precision, the position autocorrelation function, $\text{PAF(t)}=\langle x(t)x(0)\rangle$ cannot be computed directly. Rather the PAF is derived from the mean-square displacement, $\text{MSD}(t)$, via the relation $\text{PAF}(t)=k_B T/K - \text{MSD}(t)/2$. The MSD is computed from the position time trace $x(t)$ as
\begin{equation}
\text{MSD}(t)=\langle\Delta x^2(t)\rangle=\frac{1}{N}\sum_j[x(t+ t_j)-x(t_j)]^2,
\end{equation}
with $j=0, ... , N$, $t_j=j\Delta t$,  and $f_S=1/\Delta t=1\text{MHz}$ our sampling frequency.
Using the long-time limit of the MSD determined by the confinement of the trap, $\text{MSD}(t\rightarrow \infty)=2k_B T/K$, the normalized PAF reads
\begin{equation}
\text{PAF}(t)/\text{PAF}(0)=1-\text{MSD}(t)/\text{MSD}(\infty).
\end{equation}
We decomposed $x(t)$ into Fourier modes yielding the Fourier transform
\begin{equation}
\hat{x}_{\mathcal{T}}(f) = \int_{-\mathcal{T}/2}^{\mathcal{T}/2} x(t) \exp(2\pi\text{i}ft)\text{d}t,
\end{equation}
where the frequencies $f$ are integer multiples of $1/\mathcal{T}$.
The power spectral density is then obtained for long observation times as
\begin{equation}
\text{PSD}(f)=\left\langle |\hat{x}_{\mathcal{T}}(f)|^2 \right\rangle/\mathcal{T}.
\end{equation}
When plotting the PAF and the PSD on a logarithmic time, respectively, frequency scale, further data processing can be performed to extract smooth functions from noisy data$^{\text{31}}$. The data are collected in equidistant time steps and since both $\text{PSD}(f)$ and $\text{PAF}(t)$ vary essentially only on a logarithmic horizontal axis, the number of points in a log-plot will increase for long times or low frequency, respectively. Therefore, data are commonly averaged by 'blocking' consecutive data points, resulting in equidistantly distributed points on the logarithmic scale. The data scattering around their average value gives the standard error on the mean. Blocking allows to fit the data by least-square methods, which assumes that the analyzed data points are statistically independent and conform to a Gaussian distribution.

\textbf{Simulations.} The simulations are based on the method of multi-particle collision dynamics (MPC)$^{\text{26,32}}$. This mesoscale simulation technique models the solvent as a set of point particles of mass $m_\text{f}$ and density $\rho_\text{f}$ with continuous coordinates $\vec{r}$ and velocities $\vec{v}$. At every simulation step the solvent particles undergo propagation and collision. In the propagation step, the coordinates of each particle are updated according to $\vec{r}(t+\delta t) = \vec{r}(t) + \vec{v}(t)\delta t$. The particles are then sorted into collision cells, and their velocities are rotated with respect to the center-of-mass velocity of the respective collision cell by a fixed angle $\vartheta$ around a random axis. The direction of the rotation axis is the same within each collision cell, but is statistically independent between different cells and for any cell at different times. The standard grid-shifting procedure to guarantee Galilean invariance is employed.
Mass is measured in units of the solvent particle mass $m_\text{f}$, length in units of the collision cell size $b$, energy in units of the thermal energy $k_{B}T$ and time in units of $t_{0}=\sqrt{m_\text{f} b^{2}/k_{B}T}$.
Our model consists of a single colloidal particle of mass $m=335 m_\text{f}$ embedded in a sea of solvent particles with density $\rho_\text{f}=10 m_\text{f}/b^{3}$ within a simulation box with a lateral size $L=32 b$ with periodic boundary conditions. The solute-solvent particle coupling is given by a truncated Lennard-Jones potential$^{\text{33}}$ with the interaction radius $\sigma=2.0 b$. Since no angular momentum is transferred to the particle by  the collisions, we simulate effectively full-slip boundary conditions. In order to reproduce fluid-like behavior, we use  the collision time step $0.05 t_{0}$ and collision angle $\vartheta=130^{\circ}$ $^{\text{34}}$. The kinematic viscosity is then calculated$^{\text{32}}$ to be $\nu = \eta/\rho_\text{f} = 1.67 b^{2}/t_{0}$, and the Schmidt number $\text{Sc} = \nu/D= 65$.
The diffusion coefficient $D$ of the colloidal particle measured simultaneously through its mean-square displacement and the integral of the velocity auto-correlation function is $D=0.003(2) b^{2}/t_{0}$. Then, the effective hydrodynamic radius was determined to $R= 1.59b$ by the Stokes-Einstein relation $D= k_B T/\gamma$, where $\gamma = 4 \pi \eta R$ for full slip. Then the vortex diffusion time scale is calculated to $\tau_\text{f} = 1.51 t_0$, whereas the momentum relaxation time scale $\tau_\text{p} = m/\gamma$ evaluates to $\tau_\text{p} = 1.0 t_0$. The colloidal diffusion time $\tau_D = a^2/D =841 t_0$ is much larger in agreement with the experimental conditions.
Due to the high compressibility of the solvent, the sonic time $\tau_s= R/c_0 =1.59 t_0$, with the isothermal sound velocity $c_0 =\sqrt{k_B T/m_\text{f}}$, becomes comparable to vortex diffusion. Ignoring the sonic scale,
the resulting order of time scales follows the right physical hierarchy $\tau_\text{p} < \tau_\text{f} \ll \tau_D$ present in the experimental system, as prescribed in$^{\text{26}}$.
The force exerted on the colloidal particle by the optical trap is taken to be of a harmonic form, $F_\text{trap}(t)=-K x(t)$, where $K$ is the spring constant of the trap and $x(t)$ is the displacement of the particle relative to the center of the trap. The trapping potential gives rise to an additional characteristic time scale  $\tau_\text{K}=\gamma/K$.
In the simulations for parametric excitation, the spring constant of the trap is modulated as $K(t)=K [1 + g \cos(2\pi f_\text{exc} t)]$, with a relative modulation amplitude $g = 0.5$. To prevent heating, the system is thermalized with a time step $t_\text{therm}=t_{0}$ by rescaling the velocities of all particles such that the temperature of the system corresponds to the thermal energy $k_{B}T$.

\textbf{Theory.} The Langevin equation for a particle subject to Brownian motion in a harmonic potential$^{\text{22}}$ reads
\begin{equation}
m \ddot{x}(t)= F_\text{fr}(t)- K x(t) + F_\text{th}(t) ,
\end{equation}
where $x(t)$ is the displacement of the bead relative to the trap center.
The mass of the particle is again denoted by $m$, $K$ represents  the stiffness of the trap, $F_\text{fr}(t)$ the deterministic friction forces, and $F_\text{th}(t)$  is the thermal noise.
Since the hydrodynamic flow pattern has to build up first, the friction force is retarded
\begin{equation}
 F_{\text{fr}}(t) = -\int \gamma(t-t') \dot{x}(t') \text{d}t',
\end{equation}
where the kernel $\gamma(t)$ vanishes for $t<0$ by causality. By the well-known fluctuation dissipation theorem$^{\text{35}}$, the time correlation function of the random forces is determined by the same friction kernel
\begin{equation}
 \langle F_{\text{th}}(t) F_\text{th}(t') \rangle =  k_B T \gamma(|t-t'|),
\end{equation}
and $k_B T$ is the thermal energy.
It is convenient to perform a temporal Fourier transform, convention $\hat{x}(\omega) = \int \text{e}^{\text{i} \omega t} x(t) \text{d}t$, where $\omega = 2 \pi f$ denotes the angular frequency. The Langevin equation is then equivalent to
\begin{equation}
[- m \omega^2 - \text{i} \omega \hat{\gamma}(\omega)+ K ] \hat{x}(\omega) =   \hat{F}_\text{th}(\omega) .
\end{equation}
For the  correlation function of the random forces in the frequency domain one obtains
\begin{equation}
 \langle \hat{F}_{\text{th}}(\omega) \hat{F}_\text{th}(\omega')^* \rangle =  4 \pi k_B T \text{Re}[\hat{\gamma}(\omega)] \delta(\omega-\omega').
\end{equation}
For finite observation times $\mathcal{T}$, one replaces $2\pi \delta(\omega-\omega') \to\mathcal{T}$, and infers the power spectral density of the forces as $2 k_B T \text{Re}[\hat{\gamma}(\omega)]$. The colour of Brownian motion becomes manifest in the frequency dependence of the friction kernel.
Defining the Green function $\hat{G}(\omega) = [- m \omega^2 - \text{i} \omega \hat{\gamma}(\omega)+ K ]^{-1}$,
the  power spectral density (PSD) corresponding to the positional fluctuations (PAF) of the bead is  then calculated to
\begin{equation}
 \text{PSD}(\omega) = 2 k_B T \text{Re}[\gamma(\omega)] |\hat{G}(\omega)|^2 =2 k_B T \frac{ \text{Im}[\hat{G}(\omega)]}{\omega} .
\end{equation}
By the Wiener-Khinchin theorem the $\text{PAF}(t) = \langle x(t) x(0) \rangle$ is then obtained from the corresponding PSD by an inverse Fourier transform
\begin{equation}
 \text{PAF}(t) = \int \frac{\text{d}\omega}{2\pi} \text{e}^{-\text{i} \omega t}\, \text{PSD}(\omega).
\end{equation}

Einstein's theory is recovered by assuming a white noise behavior  at the frequencies of interest for the power spectral density of the forces $2k_BT\text{Re}[\hat{\gamma}(\omega)] \approx 2 k_B T{\gamma}$, with ${\gamma} = \hat{\gamma}(\omega=0)$.
 This corresponds to a time correlation function that is short lived $\langle F_{\text{fr}}(t) F_{\text{fr}}(t') \rangle = 2 k_B T {\gamma}
\delta(t-t')$ as is used in many introductory textbooks. For a spherical particle and no-slip boundary conditions the Stokes friction coefficient equals
$\gamma = 6\pi \eta R$, where $R$ is the radius of the bead.
The assumption that the friction force is given by Stokes' law is incorrect, since the long-range flow profile has to build up first. The frequency-dependent friction coefficient $\hat{\gamma}(\omega)$ corresponding to a sphere performing small-amplitude oscillations in an incompressible fluid with no-slip boundary conditions was already calculated by Stokes
\begin{equation}
 \hat{\gamma}(\omega) = 6 \pi \eta a ( 1 + \tilde{\alpha} +\tilde{\alpha}^2/9).
\end{equation}
Here $\tilde{\alpha} = \sqrt{-\text{i} \omega \tau_\text{f}}$ and the square root is to be taken with positive real part,  whereas $\tau_\text{f} = R^2 /\nu$ denotes  the vortex diffusion time.
Since $6 \pi \eta R \tau_\text{f}/9 = m_\text{f}/2$, the last term can be interpreted as effectively increasing the mass of the particle by half the mass, $m_\text{f}$, of the displaced fluid. The square root describes the evolution of the backflow by vortex diffusion and is non-analytic in frequency.
In multi-particle collision dynamics, the large compressibility of the simulated fluid gives rise to an additional time scale $\tau_\text{s} = R/c_0$, with  $c_0$ the isothermal sound velocity,  which comes close to the vortex diffusion time $\tau_\text{f}$. Furthermore, the simulations employ effectively full-slip boundary conditions, since no angular momentum is transferred to bead during the collisions with the solvent particles.
The corresponding frequency-dependent friction coefficient was calculated in$^{\text{27,36,37}}$
\begin{equation}
\hat{\gamma}(\omega) =
\frac{4 R \pi   \eta}{3} \frac{(1+\tilde{\lambda}) (18 + 18 \tilde{\alpha} + 3 \tilde{\alpha}^2 + \tilde{\alpha}^3+ 4 (1+\tilde{\alpha}) \tilde{\lambda}^2}{ (2+2 \tilde{\lambda}+\tilde{\lambda}^2) (3+\tilde{\alpha} )+ 2 (1+\tilde{\alpha} ) \tilde{\lambda}^2/\tilde{\alpha}^2}.
\end{equation}
Here $\tilde{\lambda} = R /\sqrt{ c_0^2 - \text{i} \omega (\zeta+ 4 \eta/3)/\rho_\text{f}}$ accounts for  the sound propagation, with $\zeta$ the bulk viscosity and $\rho_\text{f}$ the mass density of the fluid. One readily checks that $\hat{\gamma}(\omega=0) = 4\pi \eta R$ assumes the correct value for  the stationary friction of a sphere for full slip.\\

\begin{itemize}
\item[31.]
Flyvbjerg, H. \& Petersen, H. G. Error estimates on average of correlated data. {\it J. Chem. Phys.} {\bf91}, 461-466 (1989).
\item[32.]
Gompper, G., Ihle, T., Kroll, D. M., Winkler, R. G. {\it Advanced computer simulation approaches for soft matter sciences III}, Advances in polymer science {\bf221}, 1 (2009).
\item[33.]
Malevanets, A. \& Kapral, R. Mesoscopic Multi-particle collision model for fluid flow and molecular dynamics. Lect. Notes Phys. {\bf640}, 116-149 (2004).
\item[34.]
Ripoll, M., Mussawisade, K., Winkler, R. G., Gompper, G. Dynamic regimes of fluids simulated by multiparticle-collision dynamics. {\it Phys. Rev. E} {\bf72}, 016701 (2005).
\item[35.]
R. Kubo, M. Toda, N. Hashitsume, {\it Statistical Physics II, Nonequilibrium Statistical Mechanics}, Springer (Berlin, Heidelberg, 1991).
\item[36.]
Metiu, H., Oxtoby, D. W., Freed, K. F. Hydrodynamic theory for vibrational relaxation in liquids. {\it Phys. Rev. A} {\bf15}, 361-371 (1977).
\item[37.]
Felderhof, B. U. Effect of fluid compressibility on the flow caused by a sudden impulse applied to a sphere immersed in a viscous fluid. {\it Phys. Fluids} {\bf19}, 126101 (2007).
\end{itemize}

\end{document}